\documentclass[aps,prd,twocolumn,letterpaper,floatfix,showpacs,amsmath]{revtex4} 

\newlength{\picwidth}

\usepackage[dvips]{graphicx}

\newcommand{\ern}{\mathcal{E}}
\begin{document}

\renewcommand{\thefigure}{\arabic{figure}}
\title{

Spacetime Encodings II - Pictures of Integrability.}
\author{Jeandrew Brink}
\affiliation{
Theoretical Astrophysics, California Institute of Technology, Pasadena, CA 91103
}

\begin{abstract}

I visually explore the features of geodesic orbits in arbitrary stationary axisymmetric vacuum 
(SAV)
spacetimes that are constructed from a complex Ernst potential. Some of the geometric features of integrable and chaotic orbits are highlighted. The geodesic problem for these 
SAV
spacetimes is rewritten as a two degree of freedom problem and the connection between current ideas in dynamical systems and the study of two manifolds sought. The relationship between the Hamilton-Jacobi equations,  canonical transformations, constants of motion and Killing tensors are commented on.  Wherever possible I illustrate the concepts by means of examples from general relativity. This investigation is designed to build the readers' intuition about how integrability arises, and to summarize some of the known facts about two degree of freedom systems.
Evidence is given, in the form of orbit-crossing structure, that geodesics in SAV spacetimes might admit, a fourth constant of motion that is quartic in momentum (by contrast with Kerr spacetime, where Carter's fourth constant is quadratic).

\end{abstract}
\pacs{ }


\maketitle

\section{Introduction}
\label{sec:intro}
The Carter constant associated with a Kerr spacetime plays a crucial role in current LIGO (Laser Interferometer Gravitational Wave Observatory)   and LISA (Laser Interferometer Space Antenna) extreme and intermediate mass ratio inspiral (EMRI/ IMRI) waveform calculations. Its generalization to arbitrary stationary axisymmetric vacuum spacetimes  (SAV), could lead to an algorithm 
to map spacetimes around a compact object described by a general set of multipole moments; see \cite{FDRyan} and Paper I of
this series  \cite{JdB0} . Such an algorithm
may provide a method of determining the nature of  compact objects by asymptotically observing gravitational and electromagnetic radiation from an EMRI or IMRI inspiral.

The geodesic problem in Kerr spacetime is completely solved by the specification of four isolating integrals or constants of motion, namely rest mass, energy, axial angular momentum and the Carter constant $(\mu,\ E,\ L_z, Q)$.
The generalization of the first three constants of motion, namely $(\mu,\ E,\ L_z)$ to SAV spacetimes is trivial. These constants result from the absence of an explicit dependence of the Lagrangian on proper time, 
coordinate time and the axial angular coordinate 
respectively  $(\tau, \ t,\ \phi)$. The meaning of the fourth constant $Q$, first discovered by Brandon Carter by separation of the Hamilton-Jacobi equations  (HJE) 
\cite{CarterSeparability2,CarterSeparability}   
is a little more obscure. It is this fourth constant that allows the reduction of the geodesic equations to first-order quadratures, and the complete solution of the geodesic problem. The geometric interpretation of this fourth constant and the conditions for its existence are explored in this and subsequent papers in this series \cite{JdB2,JdB3}.

The present paper visually characterizes the geodesics in some of the axisymmetric spacetimes that are generated from an Ernst potential. In particular, the Manko-Novikov \cite{MankoNovikov1992,CastejonNovikov1990, MankoQuad} and Zipoy-Voorhees metrics \cite{Voorhees, Zipoy} are considered. The geodesic problem is formulated as a two degree of freedom problem  (2-DOF) in dynamical systems. Ideas from the field of integrable systems are collated and introduced by means of a series of visual examples. For historic purposes, the role of the Hamilton-Jacobi equation (HJE) is put in context.  Possible tests for integrability are addressed.  The concepts  of phase and energy space are introduced and illustrated by means of an example. The role and possible forms of the additional invariant are explored and a geometric interpretation of Killing tensors given.

Finally, some of the frustrations and computational difficulties when dealing with 2-DOF Hamiltonians are mentioned, and the implications of the numerical experiments in SAV spacetimes for the existence of a generalized Carter constant are described.

\section{Equations of motion}
\label{sec:EOM}
I begin with the general SAV spacetime line element of the form
\begin{align}
ds^2 &= k^2 e^{-2\psi}\left[e^{2\gamma}(d\rho^2+dz^2)+R^2d\phi^2\right]-e^{2\psi}(dt-\omega d\phi)^2, \label{LineEle}
\end{align}
where $\psi$, $\gamma$, $\omega$ and $R$ are functions of  $\rho$ and $z$ and $k$ is a real constant. The vacuum field equations relate these functions to solutions of the  Ernst equation for the complex potential~$\ern$, 
\begin{align}
\Re (\ern)\ \overline{\nabla} ^2 \ern = \overline{\nabla} \ern \cdot \overline{\nabla}\ern,
\end{align} 
where $\overline{\nabla} ^2= \partial_{\rho\rho}+\frac{1}{\rho}\partial_\rho + \partial_{zz}$, 
$\overline{\nabla} = (\partial_\rho,\partial_z)$, and the dot is the flat-space inner product.
In particular, the function $e^{2\psi} = \Re(\ern)$ denotes the real part of the potential. The functions $\gamma$ and $\omega$ can be obtained by means of line integrals of the potential once it is known. A gauge freedom in the form of the harmonic function $R$ obeying $R_{zz}+R_{\rho\rho} = 0$ exists in this metric. Often this freedom is used to set  $R=\rho$: however, for the sake of later comparison with solution generation techniques, I shall retain this generality.  

The Hamiltonian associated with geodesics of this metric is 
\begin{align}
\mathcal{H}(q,p) = \frac{1}{2}g^{\mu\nu} p_\mu p_\nu,
\end{align}
where following the notation of Goldstein \cite{Goldstein}  \mbox{$q = (\rho,\ z,\ \phi,\ t)$} are the generalized coordinates and \mbox{$p = (p_\rho,\ p_z,\ p_\phi,\ p_t)$} are the conjugate momenta. 

In order to write the equations of motion in compact form I make use of the Poisson brackets.  The Poisson bracket of two functions $g$ and $h$ with respect to the canonical variables is defined as
\begin{align}
\left[g,h\right] &= \sum_k\left(\frac{\partial g}{\partial q_k}\frac{\partial h}{\partial p_k}-\frac{\partial g}{\partial p_k}\frac{\partial h}{\partial q_k}\right).
\end{align}
The geodesic equations can now be expressed in first order form using Hamilton's equations, namely
\begin{align}
\dot{q}_{\mu} &= [q_\mu,\mathcal{H}], &\dot{p}_\mu&=[p_\mu,\mathcal{H}],
\end{align}  
where the dot $\cdot$ indicates the total derivative with respect to proper time $\tau$.

Using this notation it is immediately obvious that the absence of any explicit metric dependence on $t$ and $\phi$  results in $\dot{p}_t=\dot{p}_\phi = 0$.  By setting these quantities equal to the standard constants $p_t = -E$ and $p_\phi = L_z$, the study of geodesic motion in four-dimensional spacetime is reduced to the study of a 2-DOF dynamical system with an effective potential. The reduced Hamiltonian can be expressed as:
\begin{align}
H(\rho,z,p_\rho,p_z)= \frac{1}{2}\left(\frac{1}{V} \left(p_\rho^2+p_z^2\right) - G\right),\label{reducedH}  \end{align}
where the two potentials $V$ and $G$ have been introduced to simplify notation, and are defined as
\begin{align} 
V(\rho,z) &= k^2e^{2\gamma-2\psi},\\
G(E,L_z,\rho,z)&= -g^{AB}p_Ap_B, 
\end{align}
with $A,B$ indicating the components $t,\phi$, and let $i,j$ range over $\rho,z$.
The Hamiltonian constant $H = -1/2 \ \mu^2$  fixes the sum of the squares of the conjugate momenta to  
\begin{align}
p_\rho^2+p_z^2 = (G-\mu^2)V \equiv J(\rho,z,E,L,\mu^2), \label{Pmag}
\end{align}
and the equations of motion become
\begin{align}
\dot{q}_i &= \frac{p_i}{V},&
\dot{p}_i &= \frac{\partial_{q_i}J}{2V}.
\label{HJeq}
\end{align}
Upon introducing the non affine parameter $\lambda$ such that $V d\lambda = d\tau$, and letting $'$ indicate differentiation with respect to $\lambda$, the equations further simplify to 
\begin{align}
q_i' &= p_i,&
p_i' &= \frac{1}{2}\partial_{q_i}J.
\label{HJeq2}
\end{align}
\begin{figure}[h]
\includegraphics[width=\columnwidth]{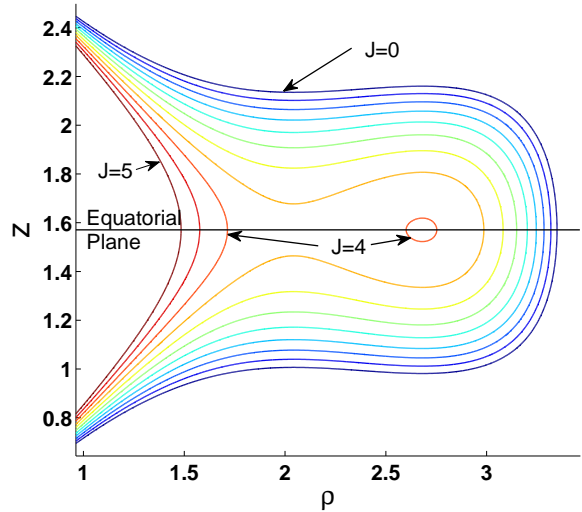}
\caption{Constant $J$ potential surfaces for a geodesic in the Schwarzschild metric with $E=0.95$, $L_z=3$, $\mu=1$. Contour Spacing 0.5. Explicit form of the metric functions can be found in Appendix \ref{MNM}}
\label{ConstVSurf}
\end{figure}
The problem of finding a generalized Carter constant in SAV spacetimes can be expressed most generally as the hunt for a function $Q(\rho,z,p_\rho,p_z)$ distinct from the Hamiltonian $H(\rho,z,p_\rho,p_z)$,  that remains constant along an orbit 
of the 2-dimensional Hamiltonian $H$, i.e.\ a function such that 
$[Q,H]=0$. Alternatively this can be stated as, the study of the geodesics of the two manifold with metric $g_J$ and associated ``Jacobi'' Hamiltonian $H_J$,
\begin{align}
{g_{J}}_{ i j} &= J \delta_{ij},& H_J &= \frac{(p_\rho^2+p_z^2)}{2J}=\frac{1}{2}. 
\end{align}
A more general and rigorous treatment of these ideas is given in \cite{Marsden}. 

In general, for a generic 
2-dimensional
Hamiltonian $H_g$, no  
such integral of motion
$Q_g$  exists and the Hamiltonian is chaotic. In most textbooks  on dynamical systems completely integrable systems are given  but  brief mention \cite{Goldstein, ChaosInt, NonlinearDynamics}. Explicit examples are rare.  A thorough review summarizing most of the known examples can be found at \cite{Hietarinta}.

\section{A Few Words about the Hamilton Jacobi Equation}
Carter's original derivation \cite{CarterSeparability2} of the fourth invariant for the Kerr metric was performed by means of separation of the Hamilton Jacobi equations (HJE) for the Jacobi function $S$, 
\begin{align}
\dot{S} = \frac{1}{2}g^{\mu\nu} \frac{\partial S}{\partial q^\mu} \frac{\partial S}{\partial q^\nu}. \label{HJEQt}
\end{align}
The Jacobi function  generates the canonical transformation to action angle variables \cite{Goldstein}(Page 449). Once $S$ is known, the problem of finding the full set of constants of motion is solved. One method of solution of \eqref{HJEQt}, and the only one so far used in practice, is by means of separation of variables. 

Lack of separation of variables, however does, not necessarily imply that the system is not integrable or the absence of $Q$. An example that does not have its origin in the vacuum field equations is the Fokas-Lagerstrom Hamiltonian \cite{Fokas}
\begin{align}
H= \frac{1}{2}(p_x^2+p_y^2)+(x^2-y^2)^{-2/3}, \label{FokasH}
\end{align}
which admits an orbital invariant
\begin{align}
Q = (p_x^2-p_y^2)(xp_y-yp_x)-4(xp_y+yp_x)(x^2-y^2)^{-2/3}.
\end{align}
No 
separation of variables for the HJE associated with \eqref{FokasH} has ever been found.

In the case of the SAV spacetimes, all metrics admitting a second-rank Killing tensor have separable HJE's  \cite{CarterSeparability, WoodhouseSep} in some coordinate system. This feature will be considered in greater detail in 
Paper III of this series
\cite{JdB2} where I will catalog the coordinate systems where this occurs.

\begin{figure}[h]
\includegraphics[width=\columnwidth]{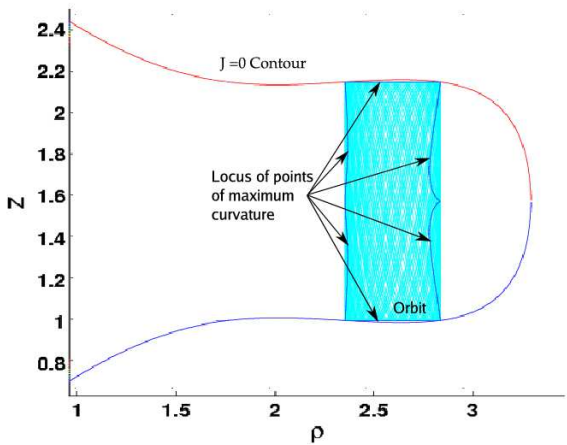}
\caption{Configuration ($\rho$, $z$) space depiction of a geodesic orbit for the potential $J$ shown in Figure \ref{ConstVSurf} (Schwarzschild metric with $E=0.95$, $L_z=3$, $\mu=1$) }
\label{OrbitJ}
\end{figure}

\section{Tests for integrability}
To my knowledge, there exist no conclusive algebraic tests for integrability for a given 2-DOF Hamiltonian $H$. Many of the difficulties in carrying out a test are summarized in  \cite{Hietarinta} and some will be demonstrated later in this paper.  Partial tests, such as the Painlev{\'{e}} test, can be carried out and all dynamical systems that pass this test have been found to be integrable. Failure to pass the Painlev{\'{e}} test however does not imply that a particular Hamiltonian will fail the test in another coordinate system. 

The Zipoy-Voorhees Metric \cite{Voorhees, Zipoy} fails the Painlev{\'{e}} test in the same manner that the Fokas-Lagerstrom Hamiltonian  expressed in the form \eqref{FokasH} does. This fact is inconclusive since after a number of difficult transformations a formulation of the Fokas-Lagerstrom Hamiltonian was found that passes the Painlev{\'{e}} test \cite{Hietarinta}.  

Possibly the strongest indication that integrability fails is the finding that numerical integration yields a Poincar{\'{e}} map without closed curves. A Poincar{\'{e}} map that displays closed curves is  indicative that the Hamiltonian may be  integrable, but it does not provide proof of the existence of an additional constant of motion.

If a Hamiltonian $H$ is close to an integrable Hamiltonian $H_0$ with invariant $Q_0$, it is always possible, following a perturbative scheme  developed by Deprit \cite{Deprit}, to compute an approximate  invariant $Q$ associated with $H$, and thus to produce an approximate Poincar{\'{e}} map with closed curves. 
A perturbative invariant so constructed, however, may not give an accurate rendition of the phase space of the perturbed Hamiltonian $H$ in a strongly chaotic regime. The classical example where this is clearly illustrated is the  H\'{e}non-Heiles problem \cite{HH}. A perturbative analysis is not sufficient to prove or disprove integrability.

Attempting to use the Deprit scheme of canonical perturbation theory to construct invariants for SAV spacetimes is prohibitively expensive computationally, and is not feasible if a solution for all SAV spacetimes is sought.

Indications of integrability can also be gleaned by observing the structure of the orbits in configuration space. This will be discussed in greater detail in  subsequent paragraphs.

In Paper IV of this series
\cite{JdB3} I will propose a test to see if SAV spacetimes admit invariants that are polynomial in momenta.

\section{Orbits, Phase Space, and Energy Space}
\begin{figure}[h]
\includegraphics[width=\columnwidth]{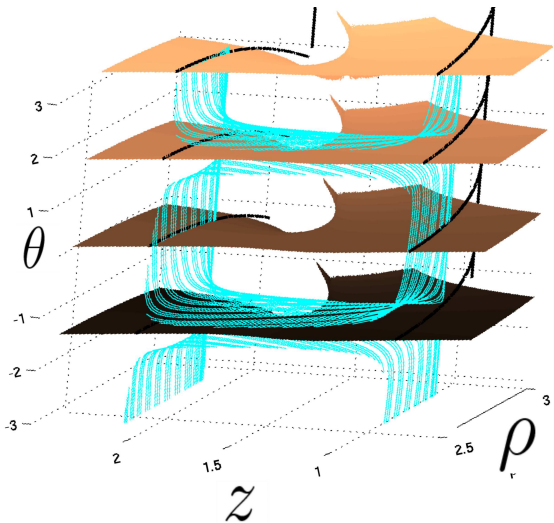}
\caption{Three dimensional energy space depiction of the torus of the orbit displayed in Figure \ref{OrbitJ}. The orbit is depicted in cyan lines. The surfaces depicted correspond  to the surfaces in phase space at which the orbit in configuration space reaches an extremum in curvature.}
\label{InvariantTorus0}
\end{figure}

Integrable systems have a surprisingly simple structure~\cite{ChaosInt}. If expressed in terms of action-angle variables, the orbits are found to trace out tori in the four dimensional phase space $(\rho,z,p_\rho,p_z)$. Although four dimensions are difficult to visualize, it is possible to see what these orbits look like in the three dimensional energy space by introducing a momentum phase angle $\theta$ such that 
\begin{align}
p_\rho &= \sqrt{J}\cos \theta, &p_z&=\sqrt{J} \sin \theta. \label{MomPhase}
\end{align}
Doing so explicitly imposes, the Hamiltonian constraint~\eqref{Pmag}, so the orbit can be visualized in ($\theta$, $\rho$, $z$) energy space, as depicted in Figure \ref{InvariantTorus0}. Note that if the light blue  lines in this figure are ``squashed'', i.e. projected down onto the ($\rho$, $z$) plane, a rotated version of Figure \ref{OrbitJ} is obtained.

One method of characterizing a curve in the $(\rho,z)$  plane (Figure \ref{OrbitJ}) that is independent of the parameterization of the curve is to compute its curvature $\kappa$. (The curvature of a curve is a measure of how rapidly the curve is moving away from its tangent line.)

The curvature of a curve parameterized by $(\rho(\tau),z(\tau))$ can be expressed as 
\begin{align}
\kappa({\tau}) &= \frac{\dot{z}\ddot{\rho}-\dot{\rho}\ddot{z}}{\left(\dot{\rho}^2+\dot{z}^2\right)^{3/2}}. \label{Curvature} 
\end{align}
Using the momentum phase angle $\theta$, this simplifies to 
\begin{align}
\kappa &= \frac{1}{2}\left(\sin \theta\ \partial_\rho (\ln J)-\cos \theta\ \partial_z(\ln J)\right). \label{eqkappa}
\end{align}
There are special points along a curve at which the curvature stops changing, or reaches an extremum, namely when \mbox{$\dot{\kappa} = 0$}.  These points are indicated by means of a dark blue line in Figure \ref{OrbitJ}.  The surfaces in energy space on which  these extrema occur can be computed for any $J$ (See appendix  \ref{ExtremeCurvatureSurfaces}).  The brown surfaces shown in Figure
\ref{InvariantTorus0} are the extreme curvature surfaces.

The rate of change of the phase angle along a particular geodesic can be calculated and expressed compactly in terms of the curvature as 
\begin{align}
\dot{\theta}&=-\frac{\kappa\sqrt{J}}{V}, &\mbox{or}&& \theta\ '&=-\kappa\sqrt{J}.  \label{thetaChange}
\end{align}

If the geodesics are integrable, as is the case in Figures~\ref{OrbitJ} and~\ref{InvariantTorus0}, the orbits sweeps out a surface in the energy space (Figure \ref{InvariantTorus0}). The locus of points at which the orbit reaches a point of extreme curvature forms a curve in configuration space (Figure \ref{OrbitJ}). The points of contact with the $J=0$ contour are unique, and are determined by the constant $Q$.  Furthermore, if a Poincare map is drawn it is constituted out of closed curves.

If the geodesic problem is not integrable (no $Q$ exists) and the orbit is strongly chaotic, it will wander all over energy space. If the geodesic problem is not integrable and the Hamiltonian $H$ is close to an integrable Hamiltonian, $H_0$, the orbit can be confined to a small volume in phase space, and appear integrable. If both $H$ and $H_0$ meet certain criteria \cite{ChaosInt}, the manner in which the surfaces in energy space, or tori in configuration space, that exist for $H_0$ are destroyed are quantified by the KAM theorem   \cite{ChaosInt}.

\section{Methods for constructing an additional invariant $Q$}
Suppose now that you suspect that an additional invariant $Q$ exists, due to a numerical exploration that yielded a Poincar{\'{e}}  map with closed curves or due to the fact that the points of extreme curvature all lie on a curve or due to your Hamiltonian passing the Painlev{\'{e}} test; and suppose you would like to construct an explicit expression for $Q$. The rules of engagement to date appear to be simple: you guess its form and hope that you are right. One method of guessing is to postulate that the invariant $Q$ is polynomial in momenta $p$.  This is equivalent to guessing that you have a Killing tensor on your two manifold. This is by no means the only form an invariant can take; however most of the known examples of integrable two dimensional Hamiltonians have polynomial $Q$'s \cite{Hietarinta}.  (For a further discussion on the generality of this form of guessing see \cite{Hall}.)

The task of finding an additional  invariant $Q$ polynomial in momenta for a 2-DOF problem in dynamical systems has a long history.  In some avenues of literature it is know as Whittaker's problem  \cite{ChaosInt,Whit}. Hall \cite{Hall} provides a very complete and readable reference and the analysis adopted here is guided largely by his treatment of the problem. This analysis will become particularly useful in a 
Paper IV
 \cite{JdB3} in this series when I return to a four dimensional representation of the geodesic problem, and attempt to understand the coupling between the Weyl tensor of SAV spacetimes and the possible existence of Killing tensors. This method of analyzing the additional invariant  appears to identify the most important quantities that should be considered, and provides a geometric picture of what they are.  Some of the difficulties in checking why a given Hamiltonian is integrable are also illustrated.  Furthermore, this approach highlights  other properties an integrable orbit has.  The problem was also considered by  \cite{Hietarinta, Xanth3}  and in different notation by \cite{Winter, Kalnins, Koenigs, Walker}.

Before I begin the analysis it is useful to introduce the complex variable  $\zeta = 1/2(\rho+iz)$. Let $\overline{\zeta}= 1/2(\rho-iz)$ denote it's complex conjugate.  In terms of this complex variable the orbital curvature in the $(\rho,z)$ plane can be expressed as 
\begin{align}
\kappa = \frac{1}{4i}\left(e^{i\theta}\partial_\zeta-e^{-i\theta} \partial_{\overline{\zeta}}\right)\ln J,
\end{align}
and derivatives along the geodesic parameterized by $\lambda$ become
\begin{align}
\partial_\lambda  &=  p_\rho\partial_\rho+p_z \partial_z = \frac{\sqrt{J}}{2}\left(e^{i\theta}\partial_\zeta+e^{-i\theta}\partial_{\overline{\zeta}}\right). 
\end{align}
Let us exploit the phase angle introduced in Eq. \eqref{MomPhase} to express our additional invariant $Q$, which is assumed to be a general $N$th order polynomial in the momenta, $p_\rho$ and $p_z$, as [cf. Eq. \eqref{MomPhase}]
\begin{align}
Q(\theta,\zeta,\overline{\zeta}   )&= \frac{1}{2} \sum_{n=-N}^{N} Q_ne^{in\theta},\label{QEQ}
\end{align}
where the $Q_n$ are complex valued functions of the configuration space variables, $Q_n=Q_n(\zeta, \overline{\zeta})$, $n$ is a positive integer and  $Q_{-n} = \overline{Q}_n$.
In effect, we are building up a Fourier series representation of the surface the orbit sweeps out in energy space, Figure \ref{InvariantTorus0}.

The condition that $Q$ is  invariant along the orbit, in other words that $Q'=0$, results in differential equations for the functions $Q_n$. Explicitly, computing $\partial_{\lambda} \eqref{QEQ}$ and making use of $\theta'= -\kappa \sqrt{J}$ yields,  
\begin{align}
Q' 
&=  \frac{\sqrt{J}}{4} \sum_{n=-N}^{N}J^{n/2} \partial_\zeta( Q_{n}J^{-n/2})  e^{i(n+1)\theta}\notag\\
& +   \frac{\sqrt{J}}{4} \sum_{n=-N}^{N} J^{-n/2}\partial_{\overline{\zeta}}(Q_nJ^{n/2} )e^{i(n-1)\theta}. 
\end{align}
If this expression is to hold for all $\theta$, the coefficients of $e^{ik\theta}$ must vanish for all $-(N+1)<k<(N+1)$, which translates into the conditions, 
\begin{align}
 \partial_\zeta  \left(\frac{ Q_n}{J^{n/2}}\right)&=0, &\mbox{for}\   n = N,N-1,\notag\\  
\partial_{\overline{\zeta}}\left(Q_{n+1}J^{\frac{n+1}{2}}\right) &=- J^{n} \partial_{\zeta}\left(\frac{Q_{n-1}}{J^{\frac{(n-1)}{2}}}\right),&  \mbox{for}\ 0<n\leq N-1,   \label{intCond}
\end{align}
where $Q_0$ is real and  $\Re(\partial_{\overline{\zeta}}(Q_1 \sqrt{J}))=0$. For negative $n$ values we get the complex conjugates of the above expressions. 
These equations can be identified directly with the Killing equations for a two-manifold. The correspondence is shown in detail in  Appendix  \ref{APPENDKT}. The lessons learned here will be exploited in that setting in my 
Paper IV
 \cite{JdB3}.

Just as in a Fourier decomposition, the equations for  odd and even $n$ decouple. Furthermore, the first condition of \eqref{intCond} implies that 
\begin{align}
\frac{Q_N}{\sqrt{J}^N}&\equiv q_N(\overline{\zeta}) \label{analF}
\end{align}
is an analytic function of $\overline{\zeta}$ and indicates an inherent ``gauge'' freedom in the Killing equations. It is this freedom that makes the identification of integrable 2-DOF Hamiltonians so difficult. The integrability conditions that the functions $Q_n$ exist result in conditions on the conformal factor $J$.
 As a result, you can write down the differential equations  the conformal factor $J$ must obey if it is to admit a Killing tensor in some coordinate system. However if you are given a sample Hamiltonian to check for integrability, you have no idea what transformation leads to the coordinate system where we can conduct the check. An additional difficulty is that the conditions on the conformal factor for $N>2$ are highly nonlinear.

In the case of SAV spacetimes that admits a second-rank Killing tensor (Carter spacetimes) it is possible to exploit the coordinate freedom to our advantage by coupling it to the gauge freedom in the metric (The R function in equation \eqref{LineEle}). A derivation of Carter spacetimes using this method is given in 
Paper III of this series
\cite{JdB2}.

The case where $N=1$, i.e. where the invariant $Q$ is linear in the momenta, corresponds to a two-metric with conformal factor $J$ that admits a Killing vector. This implies that it is a manifold of constant curvature (of the two manifold, not the orbit). The curvature of the two manifold is given by
\begin{align}
K=\frac{1}{2}\partial_{\zeta\overline{\zeta}} (\ln J)
\end{align}
 The three possibilities include flat space ($K=0$), the two sphere ($K>0$) and the Lobachevskii plane ($K<0$) which can be visualized as the surface of a bugle.

The problem of an invariant quadratic in the momenta ($N=2$),  on a two manifold was solved by Koenigs \cite{Koenigs} in 1889, who distinguished four types that are closely related to the four separable coordinate systems found by Carter (a derivation is given in 
Paper III
\cite{JdB2}) and to the super-integrable systems studied by Kalnins et al \cite{Winter,Kalnins}.  The algebraic properties of two manifolds of this type are classified in \cite{Winter, Kalnins} . Koenigs provided a very accurate geometric description of what the second quadratic invariant actually represents. This geometric picture was revisited and generalized by Moser \cite{Moser1}, and clearly illustrated by Kn\"{o}rrer \cite{Quadrics} in his study of geodesic flow on an ellipsoid. The second invariant corresponds to the Hamiltonian constant on a  two manifold distinct from the first and there exists a very simple geometric construction mapping the geodesics on the one manifold to the next. 

For the $N=4$ case very few examples are known  \cite{Hietarinta, Rosquist}.  It is  my thesis that a large class, of these two manifolds are generated by SAV spacetimes. I further suggest that the quartic structure is very closely related to the algebraic structure of the Weyl tensor, and that solution generation techniques for two manifolds already exist in the form of the solution generation techniques for SAV spacetimes. A test to see whether the SAV spacetimes admit a fourth order Killing tensor is proposed in
Paper IV of this series
\cite{JdB3}. Numerical evidence that indicates that SAV spacetimes might generate two manifolds with fourth order Killing tensor is given in Section \ref{NuMEx}.

\section{Numerical Experiments and Orbital Structure}
\label{NuMEx}

\begin{figure}[h]
\includegraphics[width=\columnwidth]{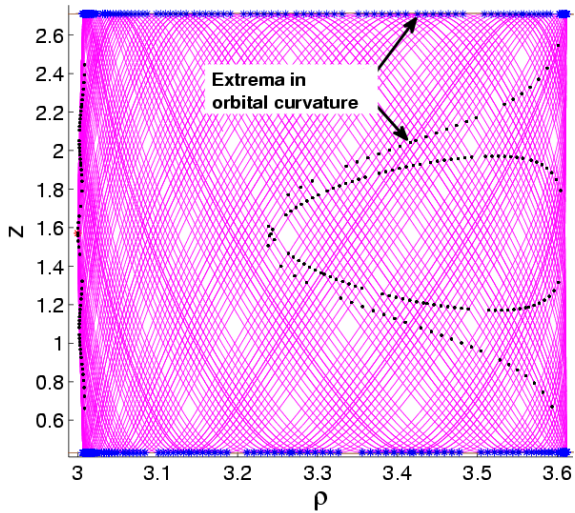}
\caption{An orbit in the Zipoy-Voorhees metric with $\delta = 2$. Orbital parameters are $E=0.95$, $L=3$, $\mu=1$.  The black dots `$\cdot$' and blue stars `*' indicate extrema in orbital curvature.  }
\label{ZVOrb}
\end{figure}

The existence of an invariant $Q$ of the form of Eq.~\eqref{QEQ}, equivalently a Killing tensor, has direct implications for the orbital appearance of a geodesic. Consider a specific point in configuration space, for example a point in Figure \ref{ZVOrb}, and consider the possible tangent directions that the geodesic could have leaving that point.
 The invariant $Q$ restricts the possible exiting tangent directions to the number of zeros of Eq.~\eqref{QEQ}.  If $N=4$ and the orbit can be traversed both ways, as in the SAV case, the answer is that there are only 4 possible crossing directions (two crossing curves). This statement must hold for every point in configuration space.  As a result, the orbits have a very ordered, cross-hatched appearance. Many numerically explored SAV spacetimes display this cross hatching pattern.

One example for which this orbital structure is observed for all parameter values I explored, is the Zipoy-Voorhees Metric  \cite{Voorhees, Zipoy}. A special case of the Weyl class, this metric has the multipole structure of a finite rod. The metric functions are given explicitly in Appendix \ref{ZVM}. It represents the one parameter ($\delta$) family of spacetimes that links flat space ($\delta =0$) to the Schwarzschild  solution ($\delta =1$).  The orbital structure is displayed in Figure \ref{ZVOrb}.  
\begin{figure}[th]
\includegraphics[width=\columnwidth]{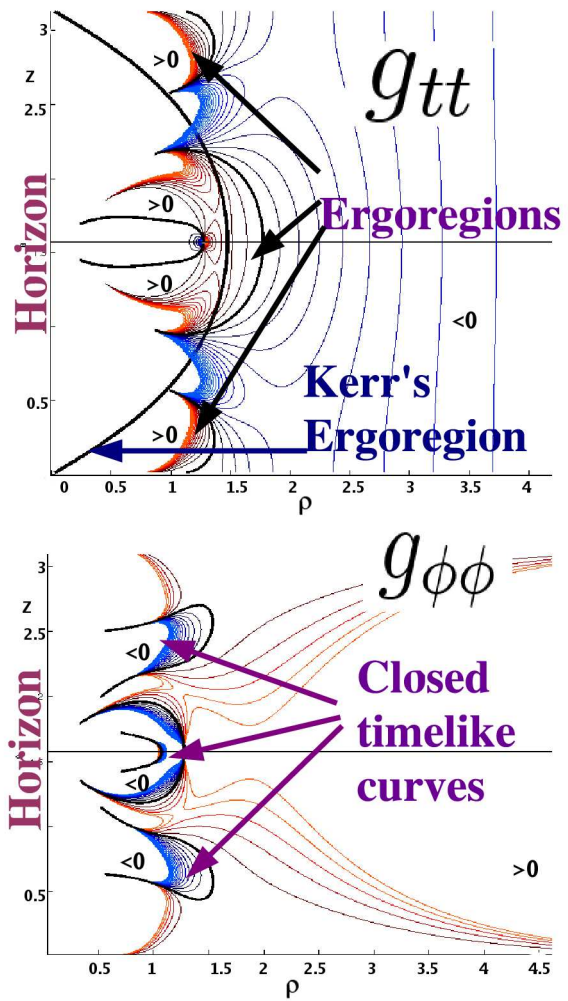}
\caption{Metric components of the Manko-Novikov spacetime. Thick black contours indicate zero values, red contours or regions marked by $>0$ admit positive values and blue contours or $<0$, negative.}
\label{ErgoManko}
\end{figure}

An example of possibly greater astrophysical application in the EMRI problem is the Manko-Novikov spacetime \cite{MankoNovikov1992,CastejonNovikov1990,MankoQuad}  whose metric components are given in Appendix~\ref{MNM}.
The initial exploration into the orbits of this spacetime was performed by Gair et al.~\cite{Gair} and unusual orbital behavior was observed.

 Some of the properties of the Manko-Novikov spacetime are sketched in Figure \ref{MankoP} and the functional form of the metric functions given in appendix \ref{MNM}.   Figure \ref{ErgoManko} characterizes the nature of the metric functions close to the horizon.
\begin{figure*}[t]
\includegraphics[width=1.75\columnwidth]{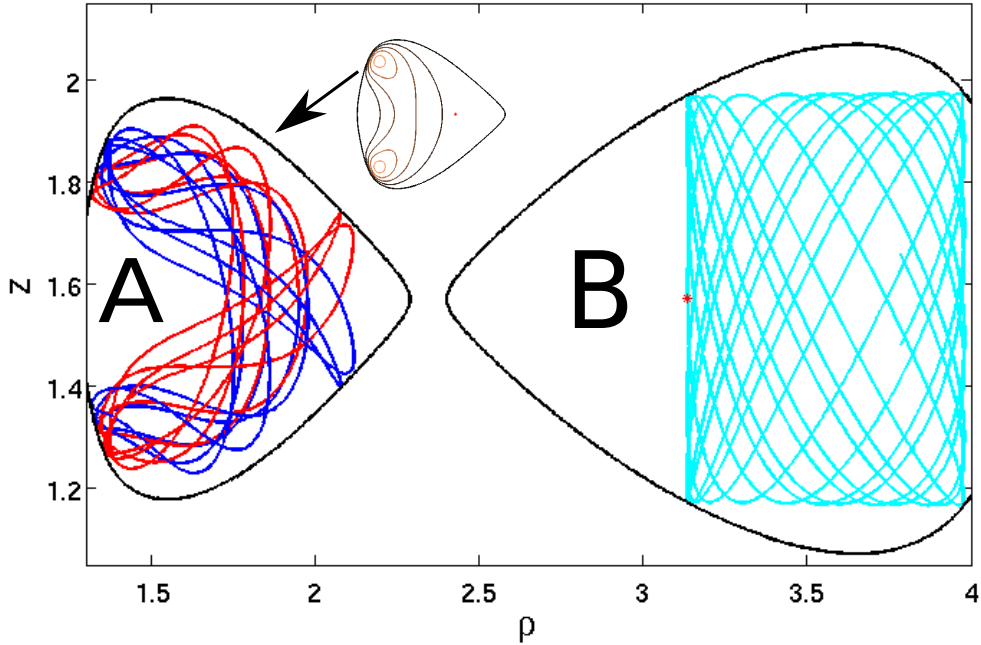}
\caption{Two orbits in the Manko-Novikov spacetime. The metric functions for this spacetime are given at the end of Appendix~\ref{MNM} and the following parameters were used $\alpha = 0.626789$,   $\alpha_2 = 11.4708$ and $k=(1-\alpha^2)/(1+\alpha^2)$. The orbital parameters associated with the orbit in the figure are $E=9.5$, $L=-3$ and $\mu =1$.}
\label{MankoOrb}
\end{figure*}

In large regions of the parameter space of this spacetime, for example region $B$ of Figure \ref{MankoOrb}, the geodesic orbits display the characteristic fourth order crossing structure. There are however regions, discovered by Gair et al. \cite{Gair}, where integrability fails and the orbit is chaotic.  One such region in which chaos occurs is displayed in Figure \ref{MankoOrb}, region A. The inset provides the contours of the potential function $J_A$ for this region;  it lies outside the ergoregions displayed in Figure \ref{ErgoManko}. The Poincar{\'{e}} maps for this orbit fail to display closed  curves \cite{Gair} and the random orbital crossing structure implies that a Killing tensor on this manifold will never be found.  I know of no similar example, in the literature of 2-DOF dynamical systems, where the orbits appear entirely integrable in one region and chaotic in another. 

The numerical experiments conducted by Gair et al. \cite{Gair} lead them to conclude that inspiralling orbits are unlikely to sample the ``chaotic'' region, so the possibility of observing such orbital behavior during a gravitational wave inspiral event is small, a conclusion with which I concur. However, conventional wisdom holds that if one observes the failure of integrability in some region of phase space it should preclude the construction of an invariant for the Hamiltonian in another. It would be an unfortunate and strange irony if ``chaotic'' behaviour in a region of phase space that is observationally inaccessible prevents us from obtaining an explicit expression for an invariant in the region of phase space from which observable gravitational radiation results.  It is this quantity that will give us theoretical power in describing inspiralling orbits in an algorithm for mapping spacetime.

Since the two regions A and B are disjoint they can be considered as two separate two manifolds $J_A$ and $J_B$, and following the analysis performed in this paper there is nothing that implies the chaos observed in region $A$ precludes the existence of a Killing tensor on the two manifold $J_B$. To date the origin of the chaos in region $A$ has not been carefully characterized. It is unclear whether the KAM theorem can be applied to this case, as the region A has no counterpart in the integrable Kerr spacetime, to which it reverts  if the anomalous multipole moments are set to zero. In many ways the explanation of the orbital behavior in region $A$  remains a very interesting puzzle.


\section{Conclusion}
This paper formulates the problem of finding  the fourth invariant,  more precisely, the generalization of Carter's constant to all SAV spacetimes, as a 2-DOF problem in dynamical systems. Equivalently stated, the problem can be formulated as the study of geodesic flow on a two manifold admitting a two metric with conformal factor $J$. A combination of the original metric functions in Eq. \eqref{LineEle} and the constants that can be easily obtained from the  metric symmetries determine $J$ (equation \eqref{Pmag}).  I summarize some related developments in dynamical systems and in the study of two manifolds, which may help this problem and point out some of the difficulties faced.  In particular, I emphasize the absence of a conclusive algebraic check of whether a two-manifold is integrable (or more specifically, possibly admits a Killing tensor), and the absence of a constructive method to construct  invariants.

The two manifold approach to the problem has the great benefit that one can visually characterize the orbits and identify the possibility of integrable behavior. It further allows one to  illustrate the  geometric meaning of a Killing tensor.   One problem faced during calculations is that the conformal factor $J$ 
is very complicated, making the complete characterization of spacetimes for a given  metric a formidable task and the characterization of all SAV spacetimes nearly impossible, using this approach.

A large class of  SAV spacetimes have orbits that appear  numerically to admit a fourth order invariant. This fact  
and the possibility of direct observational application 
if  it does (Paper I of this series  \cite{JdB0} ), 
has motivated a more in depth study
(Paper IV~\cite{JdB3}). 
It turns out that  in the context of the SAV spacetimes it may be  possible to formulate an algebraic check that will determine whether a particular spacetime admits a higher order Killing tensor and thus  quantify the relationship between the nonlocal metric distortion on a SAV spacetime described by the Weyl tensor and the dynamical behavior of  particle motion within the spacetime. This formulation however requires the full power of the tetrad formalism and the solution generation techniques, which I will review in 
Paper IV~\cite{JdB3}.

\section{Acknowledgements}
I would like to thank Frank Estabrook from whom I learnt a great deal.  I would also like to thank Ilya Mandel, Yasushi Mino, Kip Thorne and Michele Vallisneri for many useful discussions and good advice.  I greatfully acknowledge support from the Sherman Fairchild Prize Postdoctoral Fellowship for the duration of this work.

\newpage

\appendix

\section{Surfaces of extreme orbital curvature}
\label{ExtremeCurvatureSurfaces}
\begin{figure}[h]
\includegraphics[width=\columnwidth]{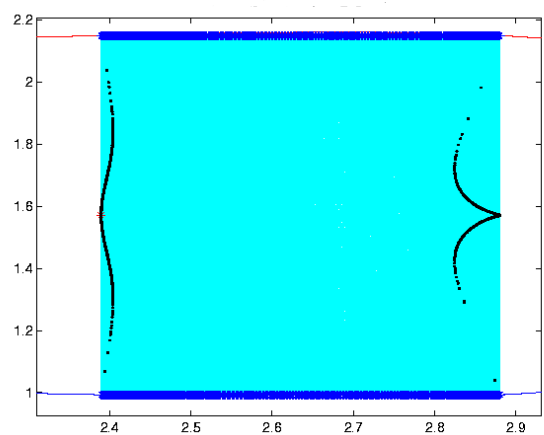}
\caption{Points of extreme orbital curvature in Schwarzschild $E=0.95$, $L=3$, $\mu=1$.
 Black dots correspond to $n=\pm 1$ surfaces.  Blue `*'  have $n=0,2$ in  equation \eqref{thetac}.
}
\label{kappado}
\end{figure}
Points where an extremum of orbital curvature is reached can be computed by setting  $\dot{\kappa}(\tau) =0$. 
After some algebra one obtains 
\begin{align}
\dot{\kappa}=[\kappa,H]= \frac{1}{2VJ^{3/2}}\left(a(\cos ^2 \theta-\sin ^2\theta)+2b\cos \theta \sin\theta\right),
\end{align}
where $a$ and $b$ are functions of $(\mu^2,E,L,\rho,z)$ defined as follows:
\begin{align}
a &= \frac{3}{2} \partial_\rho J\partial_zJ-J\partial_{\rho z}J, \notag\\
b &= \frac{1}{2}J (\partial_{\rho\rho}-\partial_{zz})J-\frac{3}{4}((\partial_\rho J)^2-(\partial_zJ)^2).
\label{abeq} 
\end{align}

As a result, curves in configuration space on which an extremum in curvature is reached  ($\dot{\kappa}=0$) can be parameterized by the phase angle $\theta_c$ at the point where
\begin{align}
\theta_c &= \frac{1}{2}\arctan\left(-\frac{a}{b}\right)+n\frac{\pi}{2}, \ \ n=\pm0,1,2. \label{thetac}
\end{align}
\begin{figure}[h]
\includegraphics[width=\columnwidth]{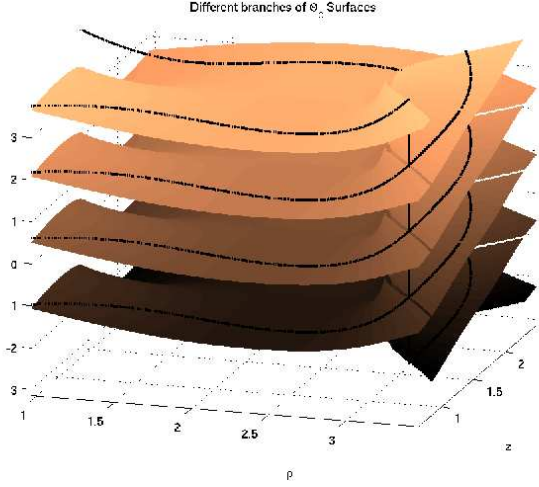}
\caption{The four branches on which solution points with extremal curvature can lie $\theta_c$ phase $E=0.95$, $L=3$, $\mu=-1$. }
\label{Brangex}
\end{figure}
The function $\theta_c(\rho,z,E,L,\mu^2)$ has four possible solution surfaces with $\theta_c \in(-\pi,\pi]$ (depicted in figure \ref{Brangex}) This function can be thought of as a phase angle surface on which all points of extremal curvature must fall. 

For each surface a branch cut occurs if $b=0$.
\begin{figure}[h]
\includegraphics[width=\columnwidth]{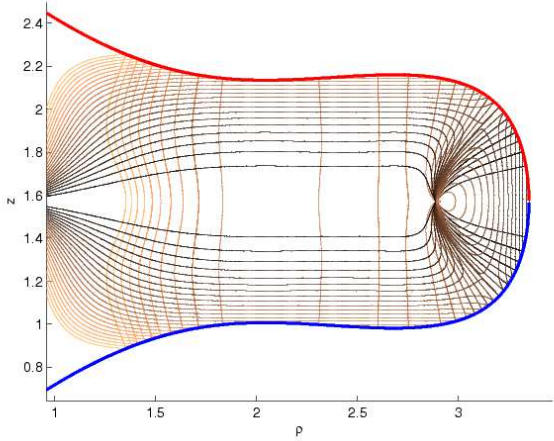}
\caption{Lines of extreme orbital curvature for several orbits of a  given $J$. Schwarzschild $E=0.95$, $L_z=3$, $\mu =1$  }
\label{EnvExtreme1}
\end{figure}
Curves of extreme orbital curvature provide an accurate way of quantifying where, given a particular $J$, a particular orbit will be confined in configuration space. In effect what is constructed is a coordinate system ideally suited to the orbits.  In Figure \ref{EnvExtreme1} the extreme orbital curvature lines for several orbits for the $J$ given in Figure \ref{OrbitJ}  are computed and the four points of contact  with the $J=0$ contour shown. Orbits on the left, that are not bound by a four pointed box, plunge through the horizon. 

 
Many quantities associated with the curvature and with extreme curvature surfaces can be most compactly expressed in complex notation. Introduce the complex variable $\zeta = 1/2(\rho+iz)$ and the complex function $c=b+ia$. Using equation \eqref{abeq}  $c$ can be expressed in terms of the the potential $J$ as follows
\begin{align}
c = \frac{1}{2}J\partial_{\zeta\zeta} J -\frac{3}{4}\left(\partial_{\zeta}J\right)^2= -J^{5/2}\partial_{\zeta\zeta}\left(\frac{1}{\sqrt{J}}\right),
\end{align}
and the extreme curvature conditions is 
\begin{align}
e^{i2\theta_C}c-e^{-2i\theta_C}\overline{c} = 0.
\end{align}

\section{Killing tensors in 2 D}
\label{APPENDKT}
The correspondence between the components of a phase space expansion $Q_n$ for calculating  the invariant $Q$ (equation \ref{QEQ}) and polynomial in momenta and components of a Killing tensor on the two manifold is given.

Consider the two metric $g_{J_{ij}}=J \delta_{ij}$ on a two manifold admitting a totally symmetric Killing tensor $T^{(\alpha_1 \cdots \alpha_N)}$ of order $N$. 
  
The Killing equations 
\begin{align}
T^{(\alpha_1 \cdots \alpha_N)}_{\ \ \ \ \ \ \ \ \ ;b} = 0, 
\end{align} 
imply that
\begin{align}
Q_T= T^{(\alpha_1 \cdots \alpha_N)}p_{\alpha_1} \cdots p_{\alpha_N} 
\end{align}
remains constant along the geodesic. Making use of the definition of the momentum phase angle, equation \eqref{MomPhase}, the invariant $Q_T$ can be rewritten as 
\begin{align}
Q_T= J^{\frac{N}{2}}  T^{(\alpha_1 \cdots \alpha_N)}\prod_{i=1}^{N} \cos \left(\theta +\frac{(1-\alpha_i)  \pi}{2} \right), 
\end{align}
where $\alpha_i=1$ indicates the index $\rho$ and $\alpha_j =2 $ the index $z$.
To put this in the form of  equation \eqref{QEQ}, let $P_{(n,N)} (i,j)$ denote the $j$-th entry of the $i$-th  permutation of a list of a total of  $N$ elements containing $N-n$ entries equal to 1 and $n$ entries equal to -1. Let $p(n)$ denote the number of permutations. As an example of the notation, let $N=4$. Then $P_{(0,4)}(1,j)=(1,1,1,1)$ has one permutation, $P_{(1,4)}$ has 4 permutations, and $P_{(2,4)}$ has 6; these are explicitely listed below 
\begin{align}
P_{(1,4)}(1,j)&=(-1,1,1,1)& P_{(1,4)}(2,j)&=(1,-1,1,1) \notag\\
P_{(1,4)}(3,j)&=(1,1,-1,1)& P_{(1,4)}(4,j)&=(1,1,1,-1)     \notag\\
P_{(2,4)}(1,j)&=(-1,-1,1,1)& P_{(2,4)}(2,j)&=(-1,1,-1,1) \notag\\
P_{(2,4)}(3,j)&=(-1,1,1,-1)& P_{(2,4)}(4,j)&=(1,-1,-1,1)     \notag\\
P_{(2,4)}(5,j)&=(1,-1,1,-1)& P_{(2,4)}(6,j)&=(1,1,-1,-1)     
 \end{align}

For clarity assume $N$ is even. (The result also holds for odd $N$ but more care has to be taken with the index ranges). The product of cosines can then be expressed as:
\begin{align}
&\prod_{i=1}^{N} \cos \left(\theta  +\beta_i \right)= \notag\\
&\frac{1}{2^N}\sum_{l=-N/2 }^{N/2} \sum_{k=1}^{p\left(\frac{N-2l}{2}\right)}  e^{2il\theta} \exp \left({\sum_{j=1}^N i\beta_j P_{(\frac{N-2l}{2} ,N )}(k,j)}\right).   
\end{align}

Thus the correspondence between the even terms in the series for even $N$ and the Killing tensor components is
\begin{align}
Q_{2l}  = 2\frac{ J^{\frac{N}{2}}}{2^N} T^{(\alpha _1 \cdots \alpha _N)} \sum_{k=1}^{p\left(\frac{N-2l}{2}\right)}   i^{ \left({\sum_{j=1}^N(1-\alpha_j)   P_{\left(\frac{N-2l}{2} ,N \right)}(k,j)}\right) }.
\end{align}
In the case where $N=2l$, the analytic function, mentioned in equation  \eqref{analF}, is 
\begin{align}
q_N(\overline{\zeta}) = \frac{Q_{N}}{\sqrt{J}^N}  = 2\frac{1 }{2^N} T^{(\alpha _1 \cdots \alpha _N)}  i^{ \left({\sum_{j=1}^N(1-\alpha_j) }\right) }.
\end{align}

The explicit expressions for the analytic function and $Q_0$ term of the lowest order case  $N=2$ are 
\begin{align}
q_2&= \frac{1}{2} (T^{\rho\rho}-T^{zz} -2 i T^{\rho z}), \notag\\
Q_0&= J (T^{\rho\rho}+T^{zz}).
\end{align}
For the $N=4$ case the expansion terms $Q_n$  expressed as a sum of the fourth order Killing tensor components are  
\begin{align}
q_4&= \frac{1}{8} (T^{\rho\rho \rho \rho} + T^{zzzz}  -6 T^{zz \rho\rho} + 4i (T^{\rho\rho\rho z} - T^{\rho z z z})),\notag\\
Q_2&= \frac{J^2}{2} \left(T^{\rho\rho\rho\rho}-T^{zzzz}-2 i (T^{\rho\rho\rho z}+ T^{\rho z z z} )\right),   \notag\\
Q_0&= \frac{3 J^2}{4} \left( T^{\rho\rho\rho\rho}+2 T^{z z \rho \rho}+  T^{z z z z} \right).   
\end{align}


\section{Manko-Novikov Metric}
\label{MNM}

\begin{figure}[h]
\includegraphics[width=\columnwidth]{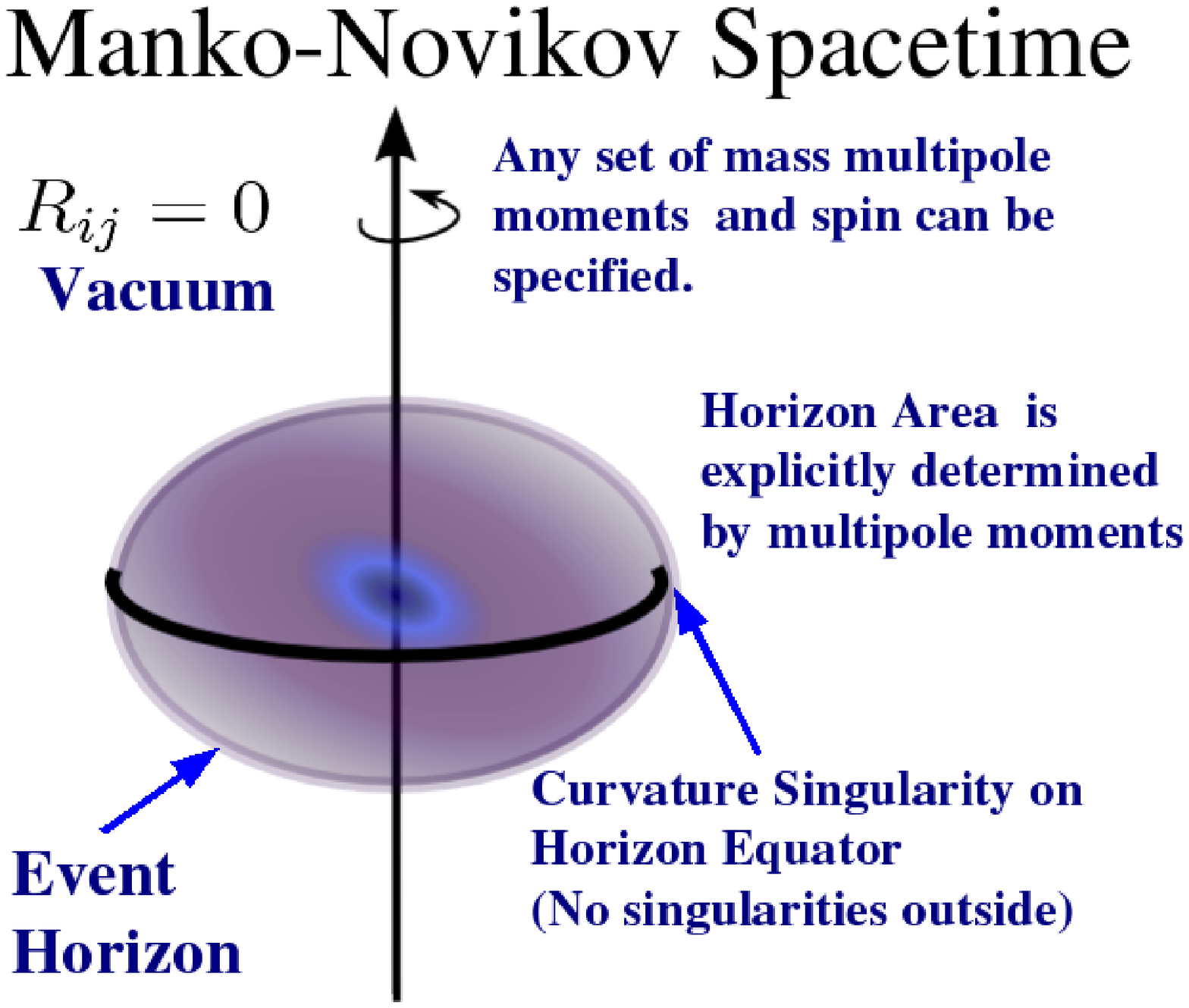}
\caption{Properties of the Manko-Novikov Spacetime.
}
\label{MankoP}
\end{figure}

The metric of the Manko-Novikov spacetime \cite{MankoNovikov1992} used to generate the plots in Figures \ref{ErgoManko}, \ref{MankoOrb} and whose properties are sketched in Figure \ref{MankoP}, can be generated from the Ernst potential of the  form $\ern = e^{2 \tilde{\psi}}A_-/A_+ $ where
\begin{align}
 A_\mp &= x(1+ab)+iy(b-a)\mp(1-ia)(1-ib),\notag\\
 \Delta \tilde{\psi} &= 0,
\end{align}
and the coordinates $x = \cosh \rho $ and $y =\cos z$ are called the Weyl coordinates. The metric functions that enter equation \eqref{LineEle} are
\begin{align}
R^2&=k^2(x^2-1)(1-y^2), &e^{2\gamma} &= e^{2\tilde{\gamma}} \frac{A  (x^2-y^2)  }{(x^2-1)(1-\alpha^2)^2},   \notag\\
e^{2\psi}   &= e^{2\tilde{\psi}} \frac{A}{B}, &
\omega &= 2k e^{-2\tilde{\psi}}\frac{C}{A}-\frac{4k \alpha}{1-\alpha^2},
\end{align}
where
\begin{align}
A &= (x^2-1)(1+ab)^2-(1-y^2)(b-a)^2,\notag\\
B &= [x+1+(x-1)ab]^2 + [(1+y)a+(1-y)b]^2,\notag\\
C &= (x^2-1)(1+ab)[b-a-y(a+b)]\notag\\
&+(1-y^2)(b-a)[1+ab+x(1-ab)]. \label{MankoNovikovgenerators}
\end{align}
The functions $a$ and $b$ obey a set of differential equations stated in \cite{MankoNovikov1992}, and one example of a solution is given below.

The solution for which the plots are made is a quadrupolar spacetime. Let $r=(x^2+y^2-1)^{1/2}$ and $u=xy/r$, and define the required Legendre polynomials to be $P_0(u)=1$, $P_1(u)=u$, $P_2(u) = -1/2+(3/2)u^2$ and $P_3(u) = u\left(5u^2-3\right)/2$. Then 
\begin{align}
\ln\left(\frac{a}{-\alpha}\right) &=-2\alpha_2\left[(x-y)\left(\frac{P_0}{r}+\frac{P_1}{r^2}+\frac{P_2}{r^3}\right) -1\right], \notag \\
\ln \left(\frac{b}{\alpha}\right) &=-2\alpha_2\left[(x+y)\left(\frac{P_0}{r} -\frac{P_1}{r^2} +\frac{P_2}{r^2}\right)-1\right], \notag\\
\tilde{\psi} &=\alpha_2\frac{ P_2}{r^3}, \notag\\
\tilde{\gamma}
&= \frac{1}{2}\ln\frac{x^2-1}{x^2-y^2}-\frac{1}{2}\left(\ln \left(\frac{a}{-\alpha}\right)+ \ln \left(\frac{b}{\alpha}\right)\right) \notag\\
&+ \alpha^2_2 \left(\frac{3}{2}\right)^2 \frac{P_3^2-P_2^2}{r^6}. 
\end{align}
Note that $\tilde{\psi}$ and $\tilde{\gamma}$ are members of the Weyl class of static metrics, $\alpha$ is a parameter that scales the spin and $\alpha_2$ is the quadrupole moment. The Geroch-Hanson multipole moments for this metric can be found in \cite{MankoNovikov1992} along with a more general solution parameterized by arbitrary mass multipole moments. In the event that $\alpha =0$ and \mbox{$\alpha_2 = 0$} the metric reduces to the Schwarzschild metric with metric functions
\begin{align}
e^{2\psi}&=\left(\frac{x-1}{x+1}\right), &
e^{2\gamma}&=\left(\frac{x^2-1}{x^2-y^2}\right), \notag\\
R^2&=(x^2-1)(1-y^2), & \omega &=0,\ \ k=1.
\end{align}

\section{Zipoy-Voorhees Metric}
\label{ZVM}
The  Zipoy-Voorhees metric \cite{Voorhees} \cite{Zipoy} is a static spacetime with metric functions
\begin{align}
e^{2\psi}&=\left(\frac{x-1}{x+1}\right)^\delta, &
e^{2\gamma}&=\left(\frac{x^2-1}{x^2-y^2}\right)^{\delta^2}, \notag\\
R^2&=(x^2-1)(1-y^2), & \omega &=0,\ \ k=1.
\end{align}
All numerical experiments performed in this metric thus far appear to admit integrable orbits similar to that portrayed in Figure 
\ref{ZVOrb}.

\bibliographystyle{apsrev}

\bibliography{BholesNemadon}

\end{document}